\documentclass[journal,10pt,letterpaper,english]{IEEEtran}
\usepackage{babel}
\usepackage[utf8x]{inputenc}
\usepackage{graphicx}
\usepackage{subfigure}
\usepackage{amsmath}
\usepackage{color}
\usepackage{balance}
\usepackage{url}
\usepackage{tikz}
\usepackage{subfigure}

\newcommand\copyrighttext{%
  \footnotesize \textcopyright 2012 IEEE. Personal use of this material is permitted.
  Permission from IEEE must be obtained for all other uses, in any current or future 
  media, including reprinting/republishing this material for advertising or promotional 
  purposes, creating new collective works, for resale or redistribution to servers or 
  lists, or reuse of any copyrighted component of this work in other works. 
  DOI: \href{https://dx.doi.org/10.1109/TCOMM.2012.022712.110142A}{10.1109/TCOMM.2012.022712.110142A}
}
\newcommand\copyrightnotice{%
\begin{tikzpicture}[remember picture,overlay]
\node[anchor=south,yshift=10pt] at (current page.south) {\fbox{\parbox{\dimexpr\textwidth-\fboxsep-\fboxrule\relax}{\copyrighttext}}};
\end{tikzpicture}%
}

\begin{document}

\title{Improving Energy Efficiency in Upstream EPON Channels by Packet
  Coalescing}
\author{Miguel~Rodríguez-Pérez,
  Sergio~Herrería-Alonso,\\
  Manuel Fernández-Veiga,~\IEEEmembership{Member,~IEEE,}
  and~Cándido~López-García%
  \thanks{The authors are with the Telematics Engineering Dept., Univ. of
    Vigo, 36310 Vigo, Spain. Tel.:+34$\,$986$\,$813458;
    fax:+34$\,$986$\,$812116; email: \protect\url{miguel@det.uvigo.es} (M.
    Rodríguez-Pérez). This work was supported by the ``Ministerio de Ciencia e
    Innovación'' through the project TEC2009-12135 of the ``Plan
    Nacional de I+D+I'' (partly financed with FEDER funds).}}

\maketitle
\copyrightnotice

\begin{abstract}
  In this paper, we research the feasibility of adapting the packet coalescing
  algorithm, used successfully in IEEE~802.3az Ethernet cards, to upstream
  EPON channels. Our simulation experiments show that, using this algorithm,
  great power savings are feasible without requiring any changes to the
  deployed access network infrastructure nor to protocols.
\end{abstract}


\section{Introduction}
\label{sec:introduction}

\IEEEPARstart{T}{he total} amount of energy needed to power networking
infrastructure has been rising quickly as more and faster devices get
connected to the Internet. In the last few years, these growing power demands
have entered in conflict with the public awareness about environmental
sustainability and the higher operating costs associated with the networking
machinery.

In this paper, we focus on reducing the power consumption of Ethernet passive
optical networks (EPONs), that are among the most popular access network
systems nowadays. Although EPONs are power efficient when operating at
$1\,$Gb/s~\cite{lange09:_energ_consum_ftth_acces_networ}, their energy needs
grow significantly when the line rate is increased to
$10\,$Gb/s~\cite{baliga09:_energ_consum_optic_ip_networ}, so an energy saving
mechanism should be employed to reduce power consumption.

Basically, an EPON consists of one optical line terminal (OLT), located at the
provider central office, connected via optical fibers and optical passive
splitters to multiple optical network units (ONUs) located at the users'
premises. In downstream transmission, data are broadcasted from the OLT to all
the connected ONUs, so each ONU must filter out those packets not directed to
itself. In the upstream direction, a common upstream channel is shared among
all the ONUs using time division multiple access (TDMA). The OLT allocates the
appropriate share of upstream bandwidth to each ONU with the help of a dynamic
bandwidth allocation (DBA) algorithm that takes into account their different
needs. Additionally, the multi-point control protocol (MPCP) is used as the
MAC algorithm to emulate a dedicated point-to-point channel from each ONU to
the OLT.

During the last years, several mechanisms to reduce the power consumption of
EPONs have appeared. Unfortunately, most of them require the modification of
the MPCP
protocol~\cite{ren10:_power_savin_mechan_perfor_analy,mandin08:_epon_power_sleep_mode,yan10:_energ_manag_mechan_ether_passiv,kubo10:_study_demon_sleep_adapt_link}
or rely on a fixed bandwidth allocation (FBA) that may result in bandwidth
under or over-allocation, since fixed time slots are assigned to each ONU
without considering their diverse bandwidth
demands~\cite{lee10:_desig_analy_novel_energ_effic}.

In this paper, we present a promising mechanism to save energy in the ONUs
allowing them to enter a low power mode when they do not need to send traffic.
We will consider that, when an ONU enters this sleep mode, its upstream
transmission circuitry can be powered down and thus, its energy demands are
minimal, though, obviously, no transmission can be carried out and any
upstream traffic must be buffered.\footnote{As we are focused in upstream
  channel savings, we consider ONUs with separate upstream/downstream
  circuitry, that can be powered down independently.} Our proposal does not
need any modification to the MPCP protocol or the DBA algorithm. We based our
work on the \emph{packet coalescing} algorithm successfully applied to reduce
power consumption in Ethernet
interfaces~\cite{christensen10:_the_road_to_eee,herreria11:_power_savin_model_burst_trans}.
Thus, instead of waking up the ONU in the presence of upstream traffic as
in~\cite{zhang11:_towar_energ_effic_epon_epon,kubo10:_study_demon_sleep_adapt_link},
we propose to delay the exit from the sleep mode until the upstream queue
reaches a certain threshold. Bringing the technique from the Ethernet domain
to EPON channels is not straightforward. The MPCP protocol constraints the ONU
restricting the transmission opportunities and demands anticipation on its
part. In this paper we propose a new state machine for sleep transitions at
the ONU that permits to use the packet coalescing algorithm even on MPCP
regulated channels. We conduct several simulation experiments to study the
effectiveness of this mechanism in terms of energy consumption and queuing
delay. Simulation results show that our proposal is able to provide great
energy savings at the expense of slightly increasing the queueing delay.

\section{Description}
\label{sec:description}

\subsection{The DBA cycle}
\label{sec:dba-cycle}

In essence, under MPCP the upstream channel is divided in periods that ONUs
employ to report their upstream queue lengths and to transmit their traffic to
the OLT. These periods correspond with DBA cycles. Each period the OLT sends a
\emph{gate message} to every ONU containing two time intervals for the next
DBA cycle: One time interval to be used to transmit data from the ONU to the
OLT and the other for sending a \emph{traffic report.} In these reports, the ONUs
indicate the amount of data stored in their upstream queues. Then, the DBA
algorithm allocates upstream capacity to each ONU based on the reports
received in the previous cycle. Thus, an ONU that had no traffic to send in
DBA cycle $i$, and hence reported empty queues to the OLT, will not receive
upstream capacity in cycle $i+1$.

When an ONU has no upstream data to transmit, it can spare to send its report
packet if the previous one was sent less than 50$\,$ms earlier. However, if
the OLT does not receive a report from an ONU for more than 50$\,$ms, the ONU
is considered disconnected by the OLT. This imposes a clear upper limit on the
time an ONU can remain continuously sleeping.

\subsection{Upstream Channel Power Saving}
\label{sec:modeling-assumptions}

\begin{figure}
  \centering
  \includegraphics[width=\columnwidth]{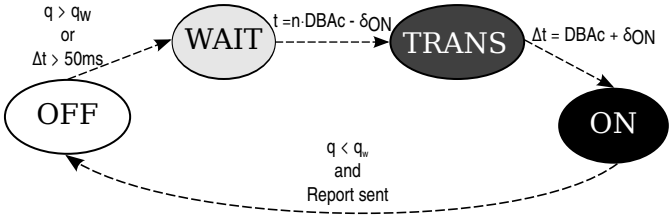}
  \caption{State machine diagram of the upstream packet coalescing
    algorithm.}
  \label{fig:state-machine}
\end{figure}
With the help of the diagram in Fig.~\ref{fig:state-machine} we will describe
the state machine of \emph{our} sleep capable ONU. Under our proposal, an ONU
remains in the OFF state until the upstream queue length $q$ surpasses a
certain threshold $q_\mathrm{w}$ or the time since the previous report was
sent reaches 50$\,$ms.
However, the ONU is not yet ready to transmit immediately after the OFF state
is left, and thus a new transient WAIT state is entered. While in WAIT, the
transmission circuitry remains off until the proper time to switch it on
arrives, as explained below.

\begin{figure}
  \centering
  \includegraphics[width=\columnwidth]{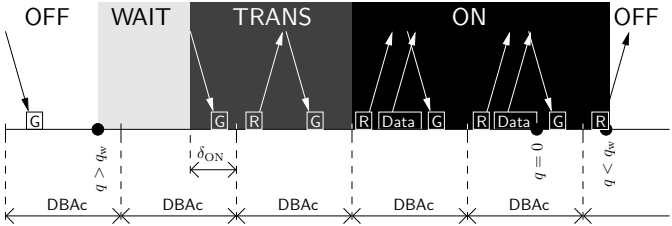}
  \caption{Time line of OFF to ON transition and back.}
  \label{fig:trantoon}
\end{figure}

Before being allowed to transmit its queued data, the ONU must first send a
report to the OLT to be allocated a transmission slot in the future DBA cycle.
However, the ONU cannot send its report immediately, but must wait until the
time indicated in the previous gate message received from the OLT. As the time
needed to power up the transmitting circuitry ($\delta_{\mathrm{ON}}$) is not
null, the ONU remains in the WAIT state until $\delta_{\mathrm{ON}}$ seconds
before the start of the next DBA cycle, in which it will be able to transmit
its report. Then, it switches to the TRANS state. Note that it is upon
entering this state that the transmission circuitry is powered on. The timing
of this procedure can also be seen in Fig.~\ref{fig:trantoon}.

The ONU stays in TRANS until it is finally allowed to start transmitting data.
For this it must wait for a whole DBA cycle. At the start of the cycle it will
report its queue length to the OLT.\footnote{The exact time will have been
  indicated in the gate message received in the previous DBA cycle.} Later in
the cycle, the OLT will send a new gate message containing both the
transmission time of the next report and the slot for data transmission, both
to occur in the next DBA cycle. So, after finishing this transitioning DBA
cycle the ONU will proceed to the \emph{normal} ON state. Note that the TRANS
state is different from the ON one in that, despite the fact that the ONU is
completely powered on in both of them, it only transmits data in ON, while in
TRANS it does no useful work. In fact, only administrative traffic (report
messages) gets transmitted.

The transition from ON to OFF is somewhat simpler. Once the ONU empties its
upstream transmission queue, it remains ON until it sends a final report in
the next DBA cycle to indicate that the queue is empty.\footnote{Another
  possibility is for the ONU to shut down immediately without sending this
  last report. Although this would increment power savings, we opted for a
  more conservative approach for the sake of interoperability.} Only when the
report is finally sent, and the upstream queue remains below the $q_\mathrm w$
threshold,\footnote{If the design opted for a more stringent approach, that
  is, demanding the queue to be completely empty when entering OFF state, the
  ONU would be prevented to save energy in some scenarios. The time between
  the end of the transmission and the final report is not negligible, and a
  few packets may arrive in between. If the ONU did not enter OFF in spite of
  those packets, it would transmit them in the next DBA cycle, defeating the
  packet coalescing strategy, and thus reducing energy efficiency.}
the ONU switches off the
transmission circuitry and enters the OFF state. We consider the time needed
to power down the circuitry to be negligible.

\section{Results}
\label{sec:results}

In this section we provide results for our proposed use of \emph{packet
  coalescing} in the upstream channel of GEPON networks. We made use of an
in-house simulator, available for download at~\cite{rodriguez11:_gepon_simul}.
For simplicity the simulator assumes a number of conditions. Mainly that the
duration of the DBA cycle is fixed and that the upstream bandwidth available
to each ONU is capped to a constant value. Upstream traffic follows a Pareto
distribution ($\alpha=2.5$). We used $200\,$Mb$/$s for the available upstream
bandwidth per ONU, as a kind of worst case scenario with fifty ONUs demanding
maximum bandwidth in a $10\,$Gb$/$s link. The rest of the experiment
parameters are detailed in Table~\ref{tab:parameters}. The transition time has
been set to the same value as
in~\cite{kubo10:_study_demon_sleep_adapt_link,mandin08:_epon_power_sleep_mode,wong09:_sleep_mode_energ_savin_pons}.
\begin{table}
  \centering
  \caption{Simulation parameters}
  \begin{tabular}{lr}
\hline
    Frame size&$1\,500\,$bytes\\
    Arrivals distribution&Pareto ($\alpha=2.5$)\\
    Upstream queue capacity&$\infty$\\
    Available upstream bandwidth&$200\,$Mb$/$s\\
    Nominal upstream bandwidth&$10\,$Gb$/$s\\
    DBA cycle length&$1.5\,$ms\\
    Active to sleep power ratio&$10:1$\\
    $\delta_{\mathrm{ON}}$&$2\,$ms\\
    $q_\mathrm w$&$1, 10 \text{ and } 100$\\\hline
  \end{tabular}
  \label{tab:parameters}
\end{table}
The power ratio between the active and sleep status
$P_{\mathrm{ON}}/P_{\mathrm{OFF}}=10$ has been chosen based on typical values
in the
literature~\cite{ren10:_power_savin_mechan_perfor_analy,mandin08:_epon_power_sleep_mode,kubo10:_study_demon_sleep_adapt_link}.\footnote{We
  consider the power consumption of the WAIT state to be equal to that of the
  OFF state, as the transmission circuitry is powered off in both states.
  Similarly, the power needs of the TRANS and ON states are also the same.}

In the first experiment we have measured the time spent in each state for
different values of the maximum queuing threshold. Each simulation was run for
$100\,$seconds and repeated with different random seeds.\footnote{$95\,\%$
  confidence intervals were negligible and are not represented so as not to
  clutter the figures.}
\begin{figure}
  \centering
  \subfigure[][$q_{\mathrm
    w}=1$\label{fig:qw-200-qw1}]{\includegraphics[width=\columnwidth]{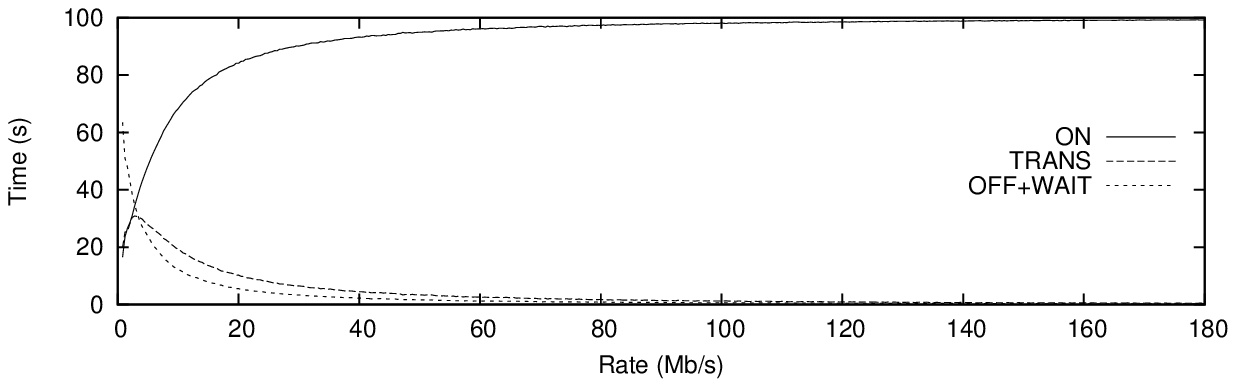}}
  \subfigure[][$q_{\mathrm
    w}=10$\label{fig:qw-200-qw10}]{\includegraphics[width=\columnwidth]{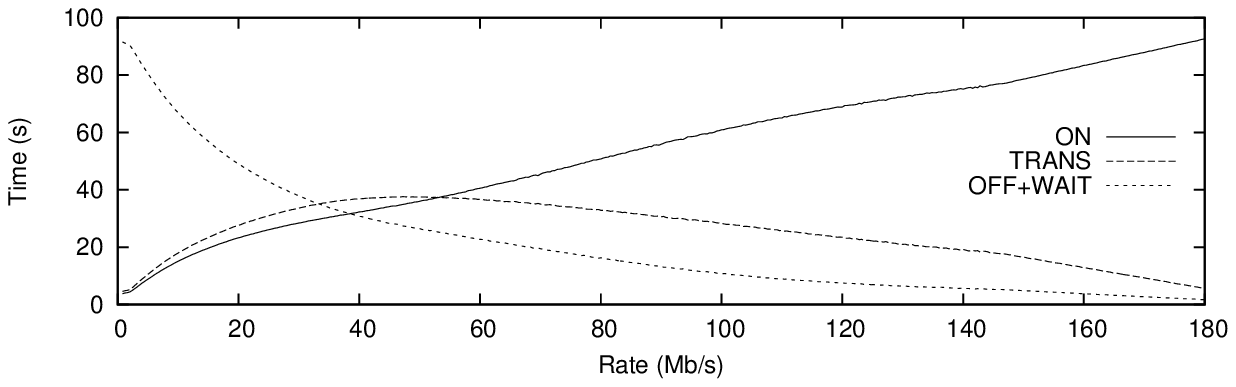}}
  \subfigure[][$q_{\mathrm w}=100$\label{fig:qw-200-qw100}]{\includegraphics[width=\columnwidth]{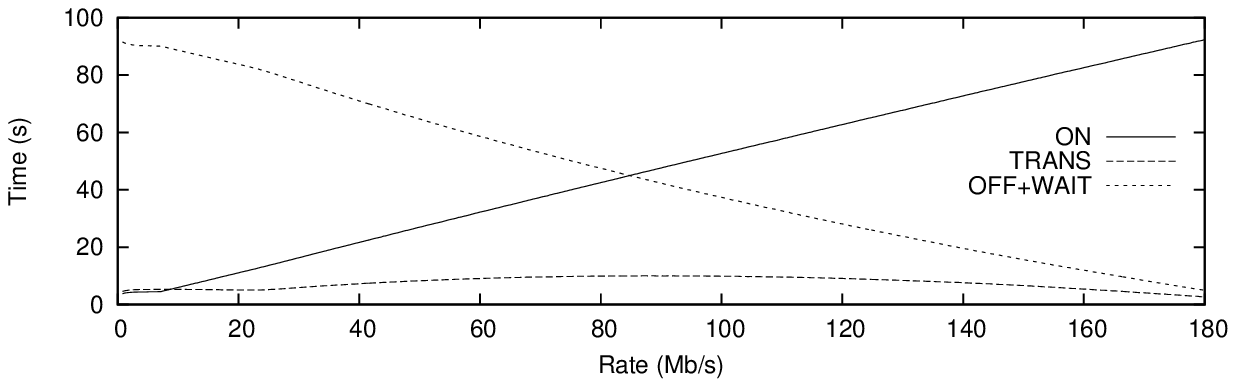}}
  \caption{Total time on each state for a $10\,$Gb/s EPON link with
    $200\,$Mb/s average upstream capacity.}
\end{figure}

Figure~\ref{fig:qw-200-qw1} shows the results obtained for $q_{\mathrm w}=1$
. As expected, the time spent ON grows with the offered load while,
conversely, the sleeping time decreases. Note, however, how the time
wasted in the TRANS state represents a significant share
specially for low loads, where a power saving algorithm should be able to save
more power. This is because, every time a new packet arrives while sleeping,
TRANS is entered wasting a fixed amount of time until the ON state is finally
reached and the backlogged traffic (a few packets, usually) is transmitted.
For moderate loads the ONU is likely ON because the probability for new
packet arrivals while waiting to send the final report and enter OFF is very
high.

If the $q_{\mathrm w}$ threshold is set to $10$ packets we obtain the results
of Fig.~\ref{fig:qw-200-qw10}. Now the time spent in ON and OFF$+$WAIT becomes
somewhat more linear. The time in TRANS still represents a significant share,
but the time ON clearly diminishes. Now there is less of a chance for $10$
packets to arrive while waiting to switch off the circuitry after a
transmission has emptied the queue.

Figure~\ref{fig:qw-200-qw100} shows the results when $q_{\mathrm w}$ is raised
to $100$ packets. This kind of extreme case shows, as expected, the best
behavior. Now the time in TRANS is very low and the ON and OFF curves are
almost linear. The TRANS time decreases because, for the same amount of
traffic, the number of needed transitions to ON is less than before, as the
ONU waits to accumulate more upstream traffic before exiting OFF. Moreover,
the time spent ON is more wisely employed, as more DBA cycles are usually
needed to transmit all the accumulated traffic. This also makes the time used
to wait for the final report less important when compared to the total time in
the ON state. That is, now the majority of the time in the ON state is
employed to actually transmit traffic.

\begin{figure}
  \centering
  \includegraphics[width=\columnwidth]{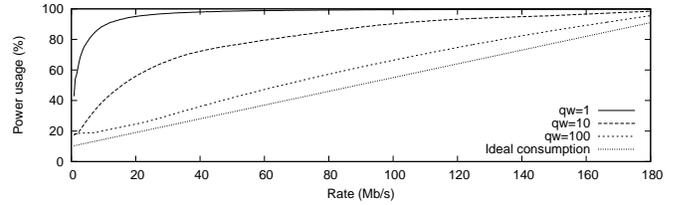}
  \caption{Power consumption in a 10$\,$Gb/s EPON link with $200\,$Mb/s
    average upstream capacity for different $q_{\mathrm w}$ thresholds.}
  \label{fig:power-200}
\end{figure}

The time spent in each status has a direct impact on the total amount of power
consumed. Figure~\ref{fig:power-200} shows the total amount of power used for
the different values of $q_{\mathrm w}$. Obviously, a non power-aware ONU
would always consume $100\,\%$ of power, while an ideal power saving mechanism
would show a linear increase in consumption. In the figure we observe how the
packet coalescing algorithm gets very good results for both $q_{\mathrm w}=10$
and $q_{\mathrm w}=100$, with the latter being really close to the ideal power
saving mechanism. With $q_{\mathrm w}=1$ however, the algorithm gets badly
punished by the frequent state transitions that waste power while not
transmitting traffic.

\begin{figure}
  \centering
  \includegraphics[width=\columnwidth]{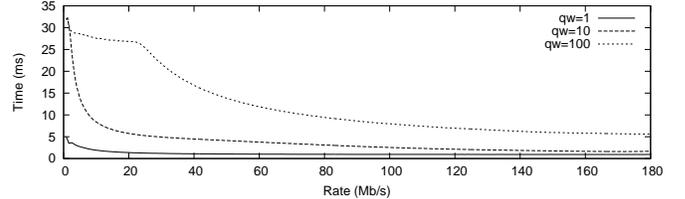}
  \caption{Average packet delay in a 10$\,$Gb/s EPON link with $200\,$Mb/s
    average upstream capacity for different $q_{\mathrm w}$ thresholds.}
  \label{fig:delay-200mbps}
\end{figure}

Our latest experiment measures the impact of the different possible
configurations of the algorithm on average packet delay. The results are
plotted in Fig.~\ref{fig:delay-200mbps}. As expected, the added delay
increases for the greatest values of $q_{\mathrm w}$. In any case, the added
delay stays always below $35\,$ms, a delay not unheard of in other last mile
technologies, like xDSL. Moreover, for low loads, where the delay is maximum,
it does not depend as much on the value of $q_{\mathrm w}$ as on the maximum
time in the OFF state. This is why both $q_{\mathrm w}=10$ and $q_{\mathrm
  w}=100$ converge for low loads. If necessary, the maximum time while
sleeping can be reduced, albeit with effects in power savings.

\section{Conclusions}
\label{sec:conclusions}

In this paper we have successfully adapted the \emph{packet coalescing}
algorithm originated in Ethernet cards to the upstream data channel of EPON
networks. The adaptation had to take into account the restrictions imposed by
the MPCP MAC protocol that regulates the access to the shared upstream channel
among the different ONUs. Simulation results show that our proposal is able to
achieve great power savings, very close in fact to the optimum, with only a
bounded moderate increase in network delay.

Finally, we are also working on coordinating the power savings in the upstream
channel with those than can be achieved if the downstream one could also be
put to sleep. This will probably require modifications to MPCP or the
cooperation of the OLT as it will have to restrict \emph{gate messages} to the
periods where the downstream interface at the ONU is ready. A mechanism to
calculate (and coordinate) optimum sleeping periods, like those present
in~\cite{gupta07:_using_low_power_modes_for}
or~\cite{rodriguez09:_improv_oppor_sleep_algor_lan_switc}, will probably be
useful.

\bibliographystyle{IEEEtran}
\bibliography{IEEEabrv,biblio}

\end{document}